# Radiative Heat Transfer and 2D Transition Metal Dichalcogenide Materials

*Long Ma, Dai-Nam Le, and Lilia M. Woods,*

Department of Physics, University of South Florida, Tampa FL 33620, USA,

Radiative heat transfer is of great interest from a fundamental point of view and for energy harvesting applications. This is a material dependent phenomenon where confined plasmonic excitations, hyperbolicity and other properties can be effective channels for enhancement, especially at the near field regime. Materials with reduced dimensions may offer further benefits of enhancement compared to the bulk systems. Here we study the radiative thermal power in the family of transition metal dichalcogenide monolayers in their H- and T-symmetries. For this purpose, the computed from first principles electronic and optical properties are then used in effective models to understand the emerging scaling laws for metals and semiconductors as well as specific materials signatures as control knobs for radiative heat transfer. Our combined approach of analytical modeling with properties from ab initio simulations can be used for other materials families to build a materials database for radiative heat transfer.

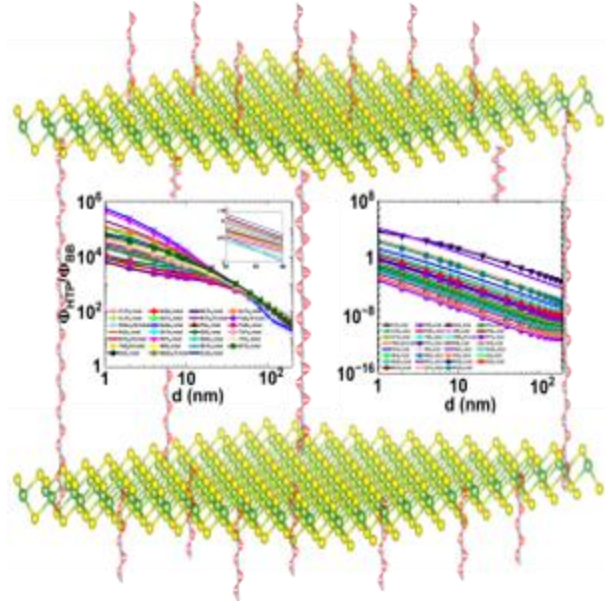

*Introduction* The omnipresent effect of radiative heat transfer between objects held at different temperatures due the exchange of fluctuating electromagnetic fields[1,2], plays a crucial role in energy harvesting applications[3,4]. In the near field regime when the objects are separated by a distance smaller than the Wien's wavelength $d < \lambda_{wien}$[5], the spatially and temporally coherent electromagnetic waves interact with each other in surface plasmon modes[6–8], giving rise to tunneled evanescent waves across the separating gap, which ultimately results in much enhanced transferred thermal power[9,10]. Such findings inspire promising applications in nano-and micro- scales for radiative heat control in near-field transistor[11], near-field grating[12], near-field solid-state cooling[13], near field thermophotovoltaic[14], and thermal scanning probe lithography[15].

Radiative power enhancement relies on optimization of the structure and material combination. Modal analysis using simplified models for the dielectric response of heat emitters taking into account their structural canonical geometries[16], hyperbolicity in uniaxial anisotropicity[17], and interpolating grating[18], gives insights into enhancing the thermal power. More realistic models utilizing the full dielectric response of the involved materials, however, can help capture various effects beyond the typical Drude or Drude-Lorentz approximations, including the underlying lattice configurations[19], electronic energy dispersion (doping)[20,21], causality and plasmonic behaviors[22] as well as scattering rates[23].

Research efforts have shown several materials with much enhanced heat transfer power. For instance, typical semiconductor-metal dual plates ($SiO_2$-Au, Si-Au) generate HTP that is hundred-fold greater than the far-field regime[24]. III-V semiconductor with tuned carrier concentrations (n,p-doped GaAs, n-doped InP, or n,p-doped and undoped InSb) exceed HTP of conventional $SiO_2$ by activating more surface plasmon channels in the infrared range[25]. Noble metals (Au, Ag, Cu) display large HTP over semiconductors due to the high reflectivity at interfaces[26]. Oxides or fluorides that support surface phonon polaritons ($Al_2O_3$ or $MgF_2$) create near-field radiation five-fold larger than the untailored $SiO_2$ at given separations[27]. Hyperbolicity in layered transition metal dichalcogenides (bulk $TiS_2$)[28] also brings promise to enhancing HTP compared to the materials without hyperbolicity. Since radiative heat transfer is a material-dependent phenomenon, efforts towards systematic studies of optimized materials for significant gain in thermal power are needed.

Recent studies have shown that 2D materials may have certain advantages for thermal radiation in the near-field regime due to stronger light-matter interactions in $MoS_2$, $MoSe_2$, $WS_2$, and $WSe_2$ monolayers when compared to their 3D counterparts[29]. 2D Dirac metals with their unique electronic structure can also experience higher order of magnitude in HTP than the bulk materials[30]. First-principles studies show that layered $\beta$-GeSe can exhibit pronounced spectral heat flux at given carrier density[31]. Metallic multilayer structures due to excited surface states give HTP that is about 40 times higher than the semi-infinite setup[32]. Considering their high efficiency in heat modulation and energy harness at specific subwavelength scale, 2D materials are attracting research interests for near-field applications. Building datasets of optical properties of 2D materials, such as transmittance, absorbance, and reflectance, as well as the decay emission rates associated with the Purcell factor[19,33] is an important step towards a broader materials-dependent understanding of optical phenomena at the nanoscale.

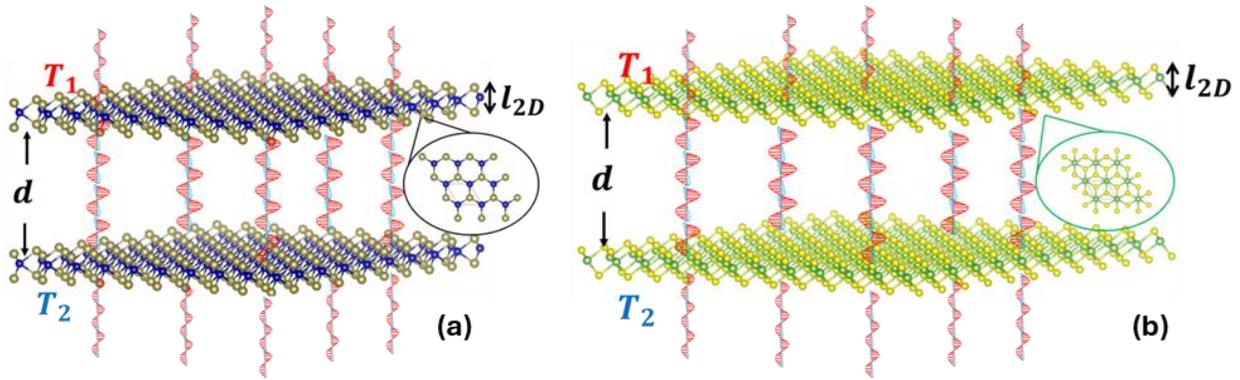

**Figure.1** Schematic representation for radiative heat transfer between planar monolayers of $AB_2$ compositions in (a) H layer group symmetry and (b) T layer group symmetry by holding the two layers at different temperature $T_1$ and $T_2$. The interlayer separation is $d$ and the monolayer thickness is $l_{2D}$.

Radiative heat transfer is a materials dependent phenomenon, thus a deeper understanding of the relation between the intrinsic properties of different composition and the thermal power is necessary. Recent studies have shown that some 2D materials have the ability to confine and control the heat radiation due to their superior electromagnetic properties. At this point, a data-driven approach focusing on a large number of nanoscaled systems is needed to broaden our materials perspective of near field transfer. In this study, we focus on the radiative heat transfer power in a family of transition metal dichalcogenide monolayered materials with trigonal $p\bar{3}m1$

(T) and hexagonal $p\bar{6}m1$ (H) symmetries. These calculations are based on the electronic and optical properties of each system obtained from first principles. The computational results are interpreted in terms of simpler analytical models which show clear relations between materials properties and radiative heat transfer control.

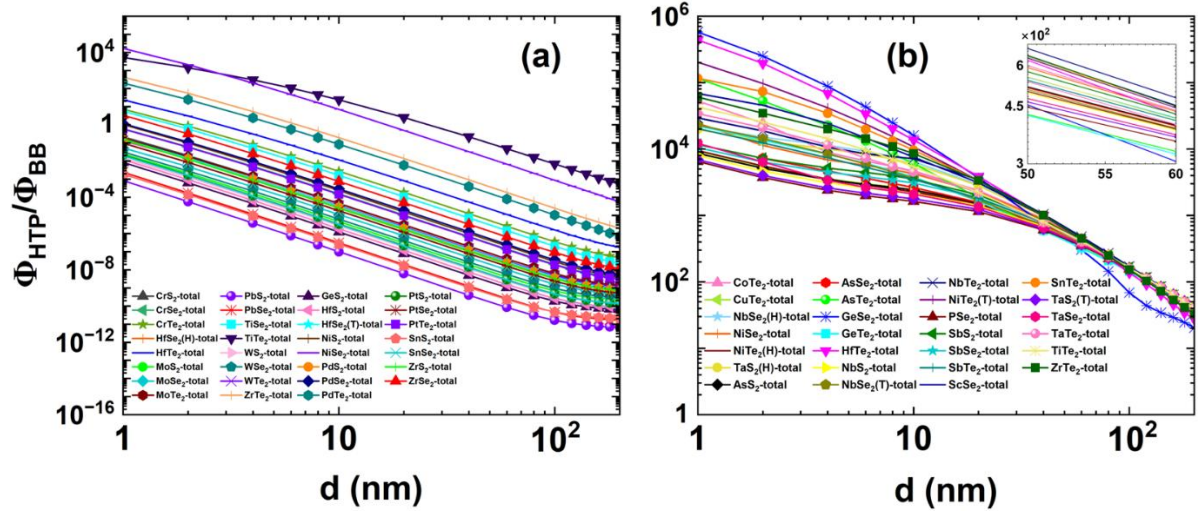

**Figure 2.** HTP normalized to the black body radiation in Stefan-Boltzmann law $\Phi_{BB} = \sigma_{SB}(T_1^4 - T_2^4)$ as a function of separation distance $d$ for (a) semiconductors and (b) metals in all symmetries. Temperatures are taken as $T_1 = 300\ K, T_2 = 400K$. The insert in panel (b) shows a blow-up of the $d = 50 \sim 60$ nm range.

*Results and Discussion* The computed optical response is then utilized to calculate HTP from Eq. 1. In Fig. 2, we show the radiative heat transfer normalized to black body radiation $\Phi_{BB} = \sigma_{SB}(T_1^4 - T_2^4)$ by giving the results for semiconductors and metals separately, where $\sigma_{SB}$ is the Stefan-Boltzmann constant. Fig. 2b shows that HTP in metals is several orders bigger than the semiconductors HTP in Fig. 2a. Each group has similar behaviors suggesting underlying commonalities, and the radiative heat transfer can vary by several orders of magnitude depending on the materials properties. For example, NiSe$_2$ (T) and TiTe$_2$ (H) exhibit the largest $\Phi_{HTP}$ among the semiconductors with $\frac{\Phi_{HTP}}{\Phi_{BB}} \sim 10^4$ for $d < 5$ nm. In comparison, HTP for PbS$_2$ (H) drops down by several orders of magnitude. For metals, GeSe$_2$ (T) outperforms the other materials especially for the distance range $d < 10$nm, where $\frac{\Phi_{HTP}}{\Phi_{BB}} \sim 10^6$. On the other hand, TaS$_2$ (T) and PSe$_2$ (T) have the lowest radiative thermal power $\frac{\Phi_{HTP}}{\Phi_{BB}} \sim 10^4$ in the same distance range. Fig. 2b also shows that for $d > 50$ nm, there is little difference between $\Phi_{HTP}$ for the different metallic monolayers, which indicates that the individual properties of the materials do not influence significantly their radiative heat power capabilities.

From the results in Fig. 1a, we find that practically all 2D semiconductors experience $\Phi \sim 1/d^4$ at separations $d < 100$nm. HTP is exclusively determined by the evanescent contribution of the $p$-modes since the propagating modes result in $\Phi_{pr}$ being several orders of magnitude smaller (see Fig. S1 in supporting information (SI)). In Fig. 2a, the displayed scaling behavior is consistent with the radiative heat transfer mediated by $p$-mode contributions between identical 2D semiconductors [28]

$$\Phi^p_{ev}(d) \approx \frac{1}{d^4}\left[1 + \frac{16\pi}{d}\left(\frac{\partial \sigma^{img}}{\partial \omega}\right)_{\omega=0}\right]\frac{\pi^4 k_B^4(T_1^4 - T_2^4)}{40\hbar^3}\left[\left(\frac{\partial^2 \sigma^{real}}{\partial \omega^2}\right)_{\omega=0}\right]^2 \quad (1)$$

where $\left(\frac{\partial^2 \sigma^{real}}{\partial \omega^2}\right)_{\omega=0} = \frac{l_{2D}}{2\pi}\frac{\omega_a^2 \gamma}{\omega_0^4}$ and $\left(\frac{\partial \sigma^{img}}{\partial \omega}\right)_{\omega=0} = -\frac{l_{2D}}{4\pi}\frac{\omega_a^2}{\omega_0^2}$. The above expression is obtained based on a single oscillator Drude-Lorentz model for the monolayer optical conductivity given by $\sigma(\omega) = \frac{l_{2D}}{4\pi} \frac{\omega_a^2 \gamma}{(\omega_0^2-\omega^2)^2+\omega^2\gamma^2} \cdot \omega^2 - i \cdot \frac{l_{2D}}{4\pi} \frac{\omega_a^2(\omega_0^2-\omega^2)}{(\omega_0^2-\omega^2)^2+\omega^2\gamma^2} \cdot \omega$ (in Gaussian units) where $\omega_a$ is the Lorentzian oscillation strength, $\omega_0$ is the interband transition frequency, and $\gamma$ is the scattering rate. Since the near field radiation is mainly influenced by exchanged photonic energy at the electrostatic limit when frequency is small, the analytical expression for $\Phi_{ev}^p(d)$ is found by exploring the $\sigma(\omega \to 0)$ limit.

While Eq. 6 captures the $d^{-4}$ distance dependence, the magnitude of $\Phi_{HTP}$ varies in a wide range for the considered materials, as also noted above. For example, at $d = 10$nm, we find that $\frac{\Phi_{HTP}}{\Phi_{BB}} = (10^{-7}, 10)$ for materials in H-symmetry and $\frac{\Phi_{HTP}}{\Phi_{BB}} = (10^{-6}, 10^{-1})$ for structures in T-symmetry. Eq. 6 suggests that the magnitude is controlled primarily by the derivative of the real part of conductivities $\left(\frac{\partial^2 \sigma^{real}}{\partial \omega^2}\right)_{\omega=0}$ and to a lesser extent by the imaginary counterpart $\left(\frac{\partial \sigma^{img}}{\partial \omega}\right)_{\omega=0}$. The Drude-Lorentz model used to obtain Eq. 6 shows that the strength of the Lorentzian oscillator $\omega_a$ and its location $\omega_0$ play a crucial role in determining the HTP magnitude. Since these are materials related properties, we further examine how the underlying electronic structure for each monolayer affects the characteristic behavior.

In Fig. 3a, we show how $\frac{\Phi_{HTP}}{\Phi_{BB}}$ evolves as a function of $\left(\frac{\partial^2 \sigma^{real}}{\partial \omega^2}\right)_{\omega=0}$. Here $\frac{\partial^2 \sigma^{real}}{\partial \omega^2}$ is obtained a numerical derivative with respect to frequency from the computationally obtained optical conductivity for each material. There is a general upward trend of HTP in terms of $\left(\frac{\partial^2 \sigma^{real}}{\partial \omega^2}\right)_{\omega=0}$ which demonstrates that larger $\left(\frac{\partial^2 \sigma^{real}}{\partial \omega^2}\right)_{\omega=0}$ results in larger transfer energy power. We find that for the majority of the materials, HTP scales are very similarly to $\left[\left(\frac{\partial^2 \sigma^{real}}{\partial \omega^2}\right)_{\omega=0}\right]^2$ consistent with analytical behavior in Eq. 6.

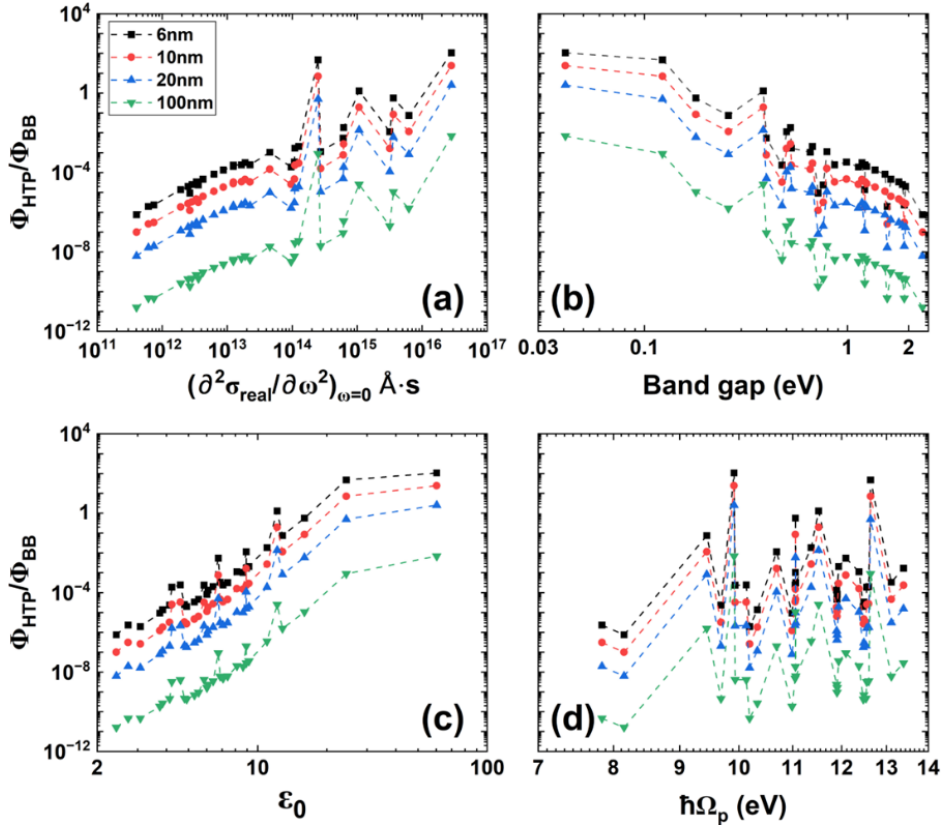

**Figure 3.** Heat transfer power for 2D semiconductors normalized to the black body limit with respect to: (a) $\left(\frac{\partial^2 \sigma^{real}}{\partial \omega^2}\right)_{\omega=0}$, (b) the energy band gap $E_g$, (c) the static dielectric constant $\varepsilon_0$, and (d) the Drude oscillator strength $\Omega_a$ at separations $d = 6nm, 10nm, 20nm, 100nm$. Each discrete point corresponds to data for a specific material, found in Table S1 in the SI file.

The Lyddane-Sachs-Teller relation[34,35] shows that the Lorentz oscillation strength $\omega_a$ and the transverse optical phonon frequency $\omega_0$ in the Drude-Lorentz model are connected with the static dielectric constant $\varepsilon_0$ and high frequency dielectric function $\varepsilon_\infty$ according to $\frac{\omega_a^2}{\omega_0^2} \sim \frac{\varepsilon_0}{\varepsilon_\infty}$. Additionally, it has been shown that for low dimensional semiconductors [36] $\varepsilon_0$ is linked with the band gap $E_g$, such that $\varepsilon_0 \sim \frac{1}{E_g^2}$. These relations indicate that the radiative heat power is correlated with $\varepsilon_0$ and $E_g$, thus in Fig. 3b,c we show how $\frac{\Phi_{HTP}}{\Phi_{BB}}$ evolves as a function of these quantities computed for each material. In particular, the overall trends suggest $\frac{\Phi_{HTP}}{\Phi_{BB}}$ decreases as $E_g$ is increased, while there is an ascending trend of HTP with $\varepsilon_0$. These heat transfer-property correlations can help us understand radiative thermal power for specific materials. For example, TiTe$_2$ in H-symmetry generates the greatest HTP among all semiconductors due to its small band gap and strong dielectric response at the small frequency range (also see Fig. S13). The corresponding parameter $\left(\frac{\partial^2 \sigma^{real}}{\partial \omega^2}\right)_{\omega=0}$ of TiTe$_2$ is the largest compared to other semiconductors reinforcing the predication by the model. In contrast, PbS$_2$ (H) has the weakest HTP which results from its smallest $\left(\frac{\partial^2 \sigma^{real}}{\partial \omega^2}\right)_{\omega=0}$ related to the large band gap and small dielectric response (also see Fig. S10).

Let us also examine how the radiative heat power changes with respect to the Lorentzian strength. Although we consider the simplest Drude-Lorentz model for Eq. 6, the computationally obtained optical response contains multiple peaks from different transitions determined by the electronic structure of the materials. From the simulations, we are able to obtain the collective Lorentzian strength from multiple peaks [23,37] according to the Thomas-Reiche-Kuhn sum rule $\Omega_p^2 = \sum_{n=1}^{N} \omega_{a,n}^2$, where $\omega_{a,n}$ is the Lorentzian oscillator at n$^{th}$ peak. In Fig. 3d, we show HTP as function of $\Omega_p$ calculated for each material. There is an oscillatory-like functionality without a clear trend between $\Phi_{HTP}$ and $\Omega_p$. This is attributed to the presence of multiple optical transitions as dictated by the electronic structure, which goes beyond the small frequency limit used for Eq. 6.

The material dependence of the heat transfer between the 2D metals can be performed in a similar way. Using a Drude-model for the optical response $\sigma(\omega) = \frac{l_{2D}\omega_p^2}{4\pi i(\omega - i\gamma)}$ where $\omega_p$ is the plasma frequency and $\gamma$ is the scattering rate, we find that for a good conductor ($\gamma \ll \omega$), $Re(\sigma) \approx \frac{\Sigma}{\omega^2}$ with $\Sigma = \frac{l_{2D}\omega_p^2\gamma}{4\pi}$. The radiative heat transfer from Eq. 1 is practically determined by the evanescent p-polarized modes, such that[28]

$$\Phi_{ev}^p(d) \approx \frac{1}{d}\left[\frac{3\,\zeta(5)k_B^5(T_1^5 - T_2^5)}{4\pi^2\hbar^4\Sigma}\right] + \frac{1}{d^2}\left[\frac{k_B^2(T_1^2 - T_2^2)}{16(48\pi^2)^{\frac{2}{3}}\hbar}e^{-\left(\frac{d_0}{d}\right)^{\frac{1}{3}}}\right], \qquad (2)$$

where $d_0 \approx \frac{\sqrt[3]{48\pi^2}\hbar^3\Sigma}{k_B^3T^3}$. The above expression shows that there is a change in the scaling law of the radiative power depending on how the distance separation compares to $d_0$. At the short separation limit ($d \ll d_0$), the exponential term in the second term approaches zero, leading to $\frac{\Phi_{HTP}}{\Phi_{BB}} \sim \frac{1}{d}$. At large separations ($d \gg d_0$), the exponential factor is close to unity giving $\frac{\Phi_{HTP}}{\Phi_{BB}} \sim \frac{1}{d^2}$. This behavior is seen in Fig. 2b, where $\frac{\Phi_{HTP}}{\Phi_{BB}} \sim \frac{1}{d}$ for $d < 10$ nm, while $\frac{\Phi_{HTP}}{\Phi_{BB}} \sim \frac{1}{d^2}$ for $d > 100$ nm. Eq.7 also tells us that the short distance limit is material dependent through the $\Sigma$ parameter, while the long-distance regime is mostly material independent. This can also be seen in Fig. 2b, where for $d > 50$ nm $\frac{\Phi_{HTP}}{\Phi_{BB}}$ converges to very similar values for practically all materials.

In Fig. 4a, we show how HTP changes as a function of $d_0 = \frac{\sqrt[3]{48\pi^2}\hbar^3\Sigma}{k_B^3 T^3}$ for the different materials, where $\Sigma$ is found computationally for each metal. For smaller separations ($d = 6$ nm, for example), there is a nonlinear decreasing trend of $\frac{\Phi_{HTP}}{\Phi_{BB}}$ as $d_0$ increases. As the separation between the layers becomes larger, this trend becomes less apparent and for $d = 100$ nm, $\frac{\Phi_{HTP}}{\Phi_{BB}}$ becomes almost independent of $d_0$.

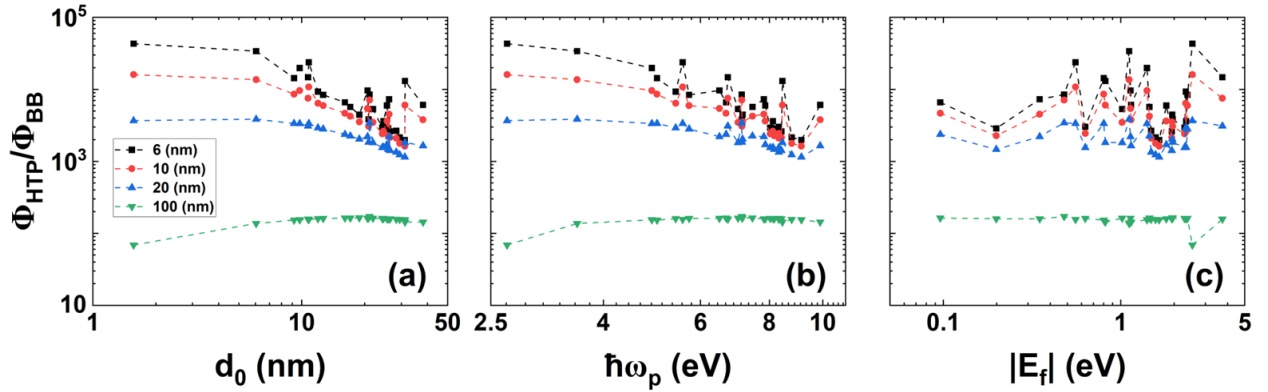

**Figure 4.** Heat transfer power for 2D metals normalized to the black body limit with respect to: (a) transition distance $d_0$ (also see SFig. 61), (b) plasma frequency $\omega_p$, and (c) Fermi energy at the separation of 6 nm, 10 nm, 20 nm, and 100 nm, respectively. The materials details are given in Table S2 in the SI file.

The materials dependence in $\Sigma$ is controlled by plasma frequency, a property directly accessible from simulations. In Fig. 4b we give $\frac{\Phi_{HTP}}{\Phi_{BB}}$ vs $\omega_p$ for each metal. A downward trend for smaller separations (shown $d = 6, 10, 20$ nm) is consistent with $\frac{\Phi_{HTP}}{\Phi_{BB}} \sim \frac{1}{\omega_p^2}$, while for $d = 100$ nm, $\frac{\Phi_{HTP}}{\Phi_{BB}}$ is almost independent of $\omega_p$. This is also consistent with the theoretical modeling from Eq. 7, showing that for $d < d_0$, $\frac{\Phi_{HTP}}{\Phi_{BB}} \sim \frac{1}{\Sigma} \sim \frac{1}{\omega_p^2}$, while for $d > d_0$, $\frac{\Phi_{HTP}}{\Phi_{BB}}$ depends very weakly on the properties of the material. These correlations can be exemplified for specific materials. For example, $\frac{\Phi_{HTP}}{\Phi_{BB}}$ of GeSe$_2$ (H) is the highest due to its smallest $\Sigma$ and $\omega_p$, while TaS$_2$ (T) and PSe$_2$ (T) generate the lowest HTP by their relatively high plasma frequency and conductivity (Table S2 in the SI file). At larger separations ($d = 100$ nm shown), all metals have similar heat transfer regardless of their properties. This can be understood by recognizing that surface plasmon excitations confined to metallic surfaces are effective channels for HTP enhancement at smaller separations. The materials dependent nature of these channels is seen in the variation of $\Phi_{HTP}$ shown in Fig. 4. At larger $d$, however, the contribution from surface plasmon is much diminished and HTP mainly results from the energy exchange of multi-reflected EM waves between the monolayers. Thus, the intensity of the radiative thermal power is determined by the size of cavity but not the materials properties.

Another property accessible from the first principles simulations is the Fermi energy $E_F$, which is closely related to the ability of metals to conduct electricity. In Fig. 4c, we show how $\frac{\Phi_{HTP}}{\Phi_{BB}}$ changes as a function of $E_F$ for the considered metallic monolayers. The behavior is highly non-monotonic without an emergent trend. There are different oscillatory like patterns that are more pronounced for shorter separations ($d =$6, 10, 20 nm shown) and for larger $d$, the heat transfer is completely independent of $E_F$.

*Conclusions* We investigate the property of radiative thermal power generated between 2D materials from the transition metal dichalcogenide family with hexagonal and trigonal symmetries.

The thermal power is calculated from the electromagnetic Poynting vector, while the electronic and optical properties for each monolayer are obtained from first principles simulations. Such an approach ensures a realistic representation of thermal radiation and the involved materials. The results for 58 monolayers H- and T-symmetry (covering 31 semiconductors and 27 metals) materials altogether are also analyzed in terms of simpler Drude and Drude-Lorentz models. The analytical expressions highlight commonalities, such as the scaling law for 2D metals and semiconductors. The $1/d^4$ distance dependence is characteristic for the semiconducting monolayers, while there is $1/d$ to $1/d^2$ transition for the metals.

Within these scaling laws, the heat transfer power can vary by several orders of magnitude depending on the particular material. The analytical modeling highlights the specific properties that are directly related to modulating the strength of thermal radiation. For semiconductors, we find that the real part of the conductivity, the band gap and dielectric constant show direct correlations with the radiative heat. In particular, systems with small energy gap and large dielectric constant exhibit larger HTP. For metals, on the other hand, the plasma frequency (which also enters in the composite Σ parameter) is perhaps the most effective control knob, especially for smaller separations in the near field regime. It is interesting to note that materials with smaller $\omega_p$ have larger HTP as opposed to those whose $\omega_p$ have much enhanced values. On the other hand, the lack of clear relation with respect to the Fermi level indicates a complex relation with the radiative thermal power.

This study shows that the realistic modeling of radiative thermal phenomena requires an integrative approach between analytical and computational calculations. While the scaling laws for metals and semiconductors are fairly robust, the heat transfer can be controlled by the properties of the materials involved. Here we have focused on the class of transition metal dichalcogenides, however, this approach can be extended by including other 2D materials such as triangular polymorph and orthorhombic metals or semiconductors with $pmmm$ and $p2mm$ symmetry[33,38]. Building a database for radiative heat transfer can broaden the materials understanding of this phenomenon and the data can be used for AI/ML models for much advanced materials predictions.

*Methodology* The theoretical modeling here relies on the electronic structure of each material as obtained from density functional theory (DFT). Our DFT simulations rely on the projector augmented wave (PAW) pseudopotentials and generalized gradient approximation-Perdew Burke Ernzerhof (GGA-PBE) exchange-correlation functionals for self-consistent field calculations. The unit cells of the 2D layers are constructed by using their symmetries. Here we consider the 2D compounds with the chemical formula $AB_2$. Materials, such as $CoTe_2$, $CrS_2$, $CrSe_2$, $CrTe_2$, $CuTe_2$, $HfSe_2$, $HfTe_2$, $MoS_2$, $MoSe_2$, $MoTe_2$, $NbSe_2$, $NiSe_2$, $NiTe_2$, $PbS_2$, $PbSe_2$, $TaS_2$, $TiSe_2$, $TiTe_2$, $WS_2$, $WSe_2$, $WTe_2$, $ZrTe_2$ fall into the hexagonal $p\bar{6}m1$ (H) symmetry and candidates, for instance, $AsS_2$, $AsSe_2$, $AsTe_2$, $GeS_2$, $GeSe_2$, $GeTe_2$, $HfS_2$, $HfSe_2$, $HfTe_2$, $NbS_2$, $NbSe_2$, $NbTe_2$, $NiS_2$, $NiSe_2$, $NiTe_2$, $PdS_2$, $PdSe_2$, $PdTe_2$, $PSe_2$, $PtS_2$, $PtSe_2$, $PtTe_2$, $SbS_2$, $SbSe_2$, $SbTe_2$, $ScSe_2$, $SnS_2$, $SnSe_2$, $SnTe_2$, $TaS_2$, $TaSe_2$, $TaTe_2$, $TiTe_2$, $ZrS_2$, $ZrSe_2$, $ZrTe_2$ with the trigonal $p\bar{3}m1$ (T) symmetry. To avoid inter-layer interactions, the materials are separated by ~20Å layers arranged in a 3D framework. The imposed relaxation criteria are $10^{-8}$ eV for energy conservation and 0.0015 eV Å$^{-1}$ for force conservation. The Gaussian smearing method with default smearing width 0.05 is chosen for all considered systems.

The dielectric function components $\boldsymbol{\varepsilon} = \text{diag}\left(\varepsilon_\parallel(\omega), \varepsilon_\parallel(\omega), \varepsilon_\perp(\omega)\right)$ are computed using the VASP and GPAW code[39] codes. We find that the VASP code renders better results for $\boldsymbol{\varepsilon}$ of semiconductors due to more accurate band structures. On the other hand, the GPAW code is better suited for metals since it allows a uniform real-space grid scheme especially for small frequencies as part of the utilized Linear Combination of Atomic Orbitals (LCAO) approach. In both scenarios, the linear dielectric response $\chi_0$ is found based on the ground state non-interacting susceptibility in which the Fourier coefficients of $\chi_0$, expanded in Bloch functions, are given in the Adler-Wiser method[40,41]. The interacting dielectric response functions are then solved from the Dyson's equation in the random field approximation (RPA) by considering local field effects and long-wavelength limit. The imaginary and real dielectric function components are related by using the Kramers-Kronig transformation. Relaxation scattering time is selected to be 10meV for all candidates. Monkhorst-Pack KPOINTS density is 10 $\text{Å}^{-1}$ used in GPAW, which is almost equivalent to $12 \times 12 \times 1$ applied in VASP scheme. The convergence of the KPOINTS is also shown in Fig. S60 for MoSe$_2$ (semiconductor) and TaS$_2$ (metal) calculated by using VASP and GPAW code.

Thermal radiation between two planar 2D monolayers held at different temperatures can be theoretically described by using fluctuation-dissipation theory. The thermally induced current and the electromagnetic field effect are framed under Maxwell's relations[6,42,43]. Provided the separation distance ($d$) and the temperature difference ($T_1$ and $T_2$) for each layer, the Heat Transfer Power ($\Phi_{HTP}$) is calculated from the ensemble average of the Poynting vector[28], yielding

$$\Phi_{HTP}(d) = \int_0^\infty \frac{d\omega}{2\pi} \Delta\Pi(\omega, T) \int_0^\infty \frac{d^2\boldsymbol{q}}{(2\pi)^2} \xi(\omega, \boldsymbol{q}, d) \tag{3}$$

where $\Delta\Pi(\omega, T)$ is the difference between Bose-Einstein distribution functions $\Pi(\omega, T_{1,2}) = \hbar\omega(e^{\hbar\omega/k_B T_{1,2}} - 1)^{-1}$, which statistically describes the net mean energy across the plates. In the spectral function $\xi(\omega, \boldsymbol{q}, d)$, the separation $d$ between the monolayers is shown in Fig. 1 and $\boldsymbol{q}$ is the surface component of wave vector $\boldsymbol{k} = (\boldsymbol{q}, k_z)$ with its normal component $k_z = \sqrt{\omega^2/c^2 - q^2}$ ($c$ is the speed of light). The relative magnitude of $|\boldsymbol{q}|$ with respect to $\omega/c$ differentiates the wave evanescence ($ev$) and propagation ($pr$) modes in the spectral function, and these contributions can be formulated as [44],

$$\xi(\omega, \boldsymbol{q}, d) = \begin{cases} Tr[(\mathbb{I} - \mathbb{R}_2^*\mathbb{R}_2 - \mathbb{T}_2^*\mathbb{T}_2)\mathbb{D}_{12}(\mathbb{I} - \mathbb{R}_1\mathbb{R}_1^* - \mathbb{T}_1^*\mathbb{T}_1)\mathbb{D}_{12}^*] & |\boldsymbol{q}| < \omega/c \\ Tr[(\mathbb{R}_2^* - \mathbb{R}_2)\mathbb{D}_{12}(\mathbb{R}_1 - \mathbb{R}_1^*)\mathbb{D}_{12}^*]e^{-2|k_z|d} & |\boldsymbol{q}| > \omega/c \end{cases} \tag{4}$$

where the Fresnel reflection matrix $\mathbb{R}_{1,2}$ is a $2 \times 2$ matrix $\mathbb{R}_{1,2} = \begin{bmatrix} R_{1,2}^{ss} & R_{1,2}^{sp} \\ R_{1,2}^{ps} & R_{1,2}^{pp} \end{bmatrix}$ and $\mathbb{D}_{12} = \frac{1}{1 - \mathbb{R}_1\mathbb{R}_2 e^{2ik_z d}}$.

For isotropic materials, the Fresnel reflection matrix is diagonal with terms corresponding to *s*- and *p*-polarized modes. It can be derived from standard electromagnetic boundary conditions for the planar geometry in Fig. 1,

$$R^{ss}_{1,2}(q,\omega) = -\frac{\frac{2\pi}{c}\sigma(\omega)}{\sqrt{1-\frac{c^2}{\omega^2}q^2}+\frac{2\pi}{c}\sigma(\omega)}, \quad R^{pp}_{1,2}(q,\omega) = \frac{\frac{2\pi}{c}\sigma(\omega)}{\frac{1}{\sqrt{1-\frac{c^2}{\omega^2}q^2}}+\frac{2\pi}{c}\sigma(\omega)}. \quad (5)$$

where $\sigma(\omega)$ is the optical conductivity of the monolayered material. With the Fresnel reflection matrix $\mathbb{R}_{1,2}$, the transmission matrix $\mathbb{T}_{1,2}$ [30,45] can be found by using $\mathbb{T}_{1,2} = \mathbb{I} - \mathbb{R}_{1,2}$.

Eqs. 1-5 establish the general framework for calculating the radiative heat transfer between isotropic 2D materials specified by their optical conductivities. In this study, we investigate a group of transition metal dichalcogenide (TMD) monolayers with the chemical formula $AB_2$ (A= Cr, Hf, Mo, Pb, Ti, W, Zr, Ge, Ni, P, Pd, Pt, Sn, Co, Cu, Nb, Ta, As, Sb, Sc, and B=S, Se, Te) by first computing their optical properties as described below. Layered group symmetries of these structures are categorized into trigonal $p\bar{3}m1$ (T) and hexagonal $p\bar{6}m1$ (H) symmetries, which takes up about ~35% of all 2D $AB_2$ configurations in the Computational 2D Materials Database (C2DB)[38]. In particular, we consider 22 monolayers in the H-symmetry and 36 monolayers in T-symmetry, which are non-magnetic and dynamically stable. Some of these TMD $AB_2$ systems were shown to have large optical reflections and strong emission properties[33]. Their 3D layered counterparts can have optical hyperbolicity with various decay rates of spontaneous emission due to tunable Purcell factors[19]. These materials may meet the demand of single photon sources over a quantum communication channel as required by quantum information technology [46].

For this study we utilize first principles simulations as implemented in the VASP and GPAW codes (See above) to compute the dielectric function components $\boldsymbol{\varepsilon} = \text{diag}\left(\varepsilon_\parallel(\omega), \varepsilon_\parallel(\omega), \varepsilon_\perp(\omega)\right)$ based on the electronic structure of the materials (See Figs. S2-S59 in supplementary file). The planar conductivity of each monolayered material is obtained as $\sigma(\omega) = \frac{(\varepsilon_\parallel(\omega)-1)\omega l_{2D}}{4\pi i}$ (in Gaussian unit), where $l_{2D}$ is the thickness of the atomic layer.

ASSOCIATED CONTENT

Information of normalized near field heat transfer power for categorized symmetries, dielectric and electronic property of each candidate material, descriptors of metals and semiconductors, convergence test for k-points mesh used in VASP and GPAW. (PDF)

AUTHOR INFORMATION

**Corresponding Author**

*Lilia M. Woods, Department of Physics, University of South Florida, Tampa FL 33620, USA. https://orcid.org/0000-0002-9872-1847; E-mail: lmwoods@usf.edu

**Authors****Corresponding Author**

*Lilia M. Woods, Department of Physics, University of South Florida, Tampa FL 33620, USA. https://orcid.org/0000-0002-9872-1847; E-mail: lmwoods@usf.edu

**Authors**

*Long Ma, Department of Physics, University of South Florida, Tampa FL 33620, USA. https://orcid.org/0000-0003-3839-2604; E-mail: longm2@usf.edu

*Dai-Nam Le, Department of Physics, University of South Florida, Tampa FL 33620, USA. https://orcid.org/0000-0003-0756-8742; E-mail: dainamle@usf.edu

**Author Contributions**

The project was conceived and designed by L.M.W. The simulations were performed by performed by L.M and the modeling was done by D.N.L. The analysis and manuscript writing was done by L.M.W and L.M. with contributions from D.N.L.

**Conflicts**

The authors declare no competing interests.

**Data Availability**

The data supporting this article (results from first principles simulations of the electronic and dielectric properties of the considered 2D materials) is included in the main text and the Supplementary Information file.

ACKNOWLEDGMENT

This work is supported by the US Department of Energy under Grant No.DE-FG02-06ER46297. Computational resources are provided by USF Research Computing.